\documentclass[aps,twocolumn,floatfix,groupedaddress,nofootinbib]{revtex4}
\usepackage{color}
\usepackage[usenames,dvipsnames,svgnames,table]{xcolor}
\usepackage{mathtools}
\usepackage{graphicx}
\usepackage{dcolumn}
\usepackage{array}
\usepackage{lipsum}
\usepackage{bm}
\usepackage{subfigure}
\usepackage{amssymb}
\usepackage{multirow}
\usepackage{tabularx}
\usepackage{amsmath}
\usepackage{braket}
\graphicspath{{plots/}}
\usepackage{lipsum}
\usepackage{mathrsfs}

\begin{document}

\title{Unravelling the nonlinear ideal density response of many-body systems}
\author{Panagiotis Tolias$^{1}\footnotemark\footnotetext{\vspace*{-2.40mm}corresponding author: tolias@kth.se\vspace*{-10.40mm}}$, Tobias Dornheim$^{2,3}$, Zhandos A. Moldabekov$^{2,3}$ and Jan Vorberger$^{3}$}
\affiliation{$^1$ Space and Plasma Physics - Royal Institute of Technology (KTH), SE-10044 Stockholm, Sweden\\
             $^2$ Center for Advanced Systems Understanding (CASUS), D-02826 G\"orlitz, Germany\\
             $^3$ Helmholtz-Zentrum Dresden-Rossendorf (HZDR), D-01328 Dresden, Germany\vspace*{-1.80mm}}
\begin{abstract}
\noindent Nonlinear density response theory is revisited focusing on the harmonically perturbed finite temperature uniform electron gas. Within the non-interacting limit, brute force quantum kinetic theory calculations for the quadratic, cubic, quartic and quintic responses reveal a deep connection with the linear response. Careful analysis of the static long wavelength limit led us to conjecture a canonical non-interacting form that expresses arbitrary order nonlinear responses as the weighted sum of the linear responses evaluated at all multiple harmonics. This harmonic expansion is successfully validated against \emph{ab initio} path integral Monte Carlo simulations.
\end{abstract}
\maketitle

\textbf{Introduction}\,-- The assumption of the linear response of a system to infinitesimal perturbations is ubiquitous in physics fields\,\cite{generalR1,generalR2,generalR3}. The associated theoretical framework, linear response theory, enables the description of static and dynamic phenomena on the basis of the equilibrium properties of the unperturbed system alone. The infinitesimal perturbation strength restriction can be relaxed in nonlinear response theory with the equilibrium connection retained and one of the caveats being an often unmanageable increase in mathematical complexity\,\cite{generalR4,generalR5}.

Nonlinear dielectric permittivities have been traditionally investigated in \emph{ideal classical plasmas}, where relatively strong electrostatic fields can develop even spontaneously, \emph{i.e.,} in absence of intense external fields, owing to the growth of a host of instabilities. Progress has been driven by interest in cosmic rays, solar activity, magnetic confinement fusion, plasma turbulence and laser plasma interactions\,\cite{introduc1,introduc2,introduc3,introduc4,introduc5,introduc6,introduc7}. These studies exclusively focused on three- and four-wave resonances, thus, they were limited to quadratic and cubic responses\,\cite{introduc2,introduc3}. Nonlinear density responses have been investigated in \emph{strongly coupled plasmas} (the classical one-component plasma -- OCP), with interest lately extending from the quadratic order\,\cite{OCPnonli1,OCPnonli2} to the cubic and quartic orders\,\cite{OCPnonli3}. The quadratic density response of the three- and two- dimensional fully degenerate \emph{uniform electron gas} (UEG) (the quantum one-component plasma -- qOCP), has also been investigated in numerous works\,\cite{OCPnonli4,OCPnonli5,OCPnonli6,OCPnonli7,Mikhailo1,Mikhailo2}, with attention mostly paid to the electronic stopping power for high-Z ions\,\cite{stopping1,stopping2,stopping3,stopping4} and the construction of effective pair potentials\,\cite{effectiv1,effectiv2}.

The accurate theoretical description of the {\em finite temperature} dense UEG is pivotal to our understanding of warm dense matter (WDM)\,\cite{DorReview,BonReview,PoPReview}, an extreme state of high temperature highly compressed matter\,\cite{WDMrefer1,WDMrefer2} relevant for dense astrophysical objects\,\cite{WDMrefer3}, inertial confinement fusion\,\cite{WDMrefer4} and novel material fabrication\,\cite{WDMrefer5}. The dimensionless parameters that define the UEG state points; \textbf{(1)} the quantum coupling parameter $r_{\mathrm{s}}=d/a_{\mathrm{B}}$ with $d=(4\pi{n}/3)^{-1/3}$ the Wigner Seitz radius, $a_{\mathrm{B}}=\hbar^2/m_{\mathrm{e}}e^2$ the Bohr radius, \textbf{(2)} the degeneracy parameter $\Theta=k_\textnormal{B}T/E_{\mathrm{F}}$ with $T$ the temperature, $E_{\mathrm{F}}=\hbar^2q_{\mathrm{F}}^2/(2m_{\mathrm{e}})$ the Fermi energy, are both of the order of unity in WDM conditions. The lack of small parameters is indicative of the difficulties in the theoretical treatment of the complex interplay between quantum effects, Coulomb correlations and thermal excitations in the WDM regime. Nevertheless, in the last decade, breakthroughs in quantum Monte Carlo simulations\,\cite{DorReview,BonReview} have led to a very accurate description of the warm dense UEG thermodynamic properties\,\cite{WDMtherm1,WDMtherm2} as well as static\,\cite{WDMstati1,WDMstati2} and dynamic behavior\,\cite{WDMdynam1,WDMdynam2,WDMdynam3} at the level of the linear response and two-body density correlations. This state-of-the-art has been pushed even further in a recent series of works, where Dornheim \& coworkers reported systematic \emph{ab initio} path integral Monte Carlo (PIMC) results for the nonlinear density response of the warm dense UEG\,\cite{Dornheim1,Dornheim2,Dornheim3,Dornheim4,Dornheim5,Dornheim6}. Very recently, thermal Kohn--Sham density functional theory (KSDFT) was also revealed to provide remarkably accurate results for the (non)linear UEG density response\,\cite{Dornheim7}.

Nonlinear effects in WDM and the UEG are an essential ingredient for the construction of effective screened ion potentials at short distances\,\cite{applicat1}, the modelling of the effective electronic interactions beyond the Kukkonen-Overhauser potential\,\cite{applicat2,applicat3}, the understanding of dynamic many-body correlations especially towards strong coupling\,\cite{applicat4,applicat5,applicat6}, the description of multi-plasmon excitations\,\cite{applicat7} and the accurate estimation of the energy-loss characteristics of high-Z ions\,\cite{applicat8}. More important, it has been concluded that nonlinear responses are much more sensitive on the thermodynamic state (\emph{e.g.} the temperature) than linear responses, which makes them a potential diagnostic tool for WDM experiments\,\cite{Dornheim1,Dornheim2,applicat9,applicat0}.

Systematic comparison of PIMC simulation results for the quadratic and cubic static density responses of the warm dense UEG with quantum statistical theory has revealed that the random phase approximation (taking into account screening on the linear response level) supplemented with a near-exact two-body static local field correction, yields excellent predictions\,\cite{Dornheim2,Dornheim3}. The main obstacle in the exploration of even higher nonlinearities is the lack of analytical expressions for the nonlinear non-interacting density responses at higher harmonics\,\cite{Dornheim7}.

In this Letter, some fundamental results are derived from a general consideration of the quantum theory of the nonlinear response. These culminate into a practical recursion formula that expresses the harmonic ideal density response as a sum of Lindhard functions evaluated at multiple harmonics. This expression is validated against novel PIMC simulations of the non-interacting UEG.

\textbf{Nonlinear density response theory}\,--\,Let $U_{\mathrm{ext}}(\boldsymbol{r},t)$ be an external potential energy perturbation that is applied to a homogeneous stationary system. The induced density ${n}_{\mathrm{ind}}(\boldsymbol{r},t)$ can be considered to be a functional of the external potential energy, ${n}_{\mathrm{ind}}[U_{\mathrm{ext}}]$\,\cite{introduc3}. The associated functional Taylor expansion can be expressed as
\begin{align*}
{n}_{\mathrm{ind}}(\boldsymbol{r},t)&=\sum_{l=1}^{\infty}\int\mathscr{W}^{(l)}(\{l\})\prod_{i=1}^{l}U_{\mathrm{ext}}(i)d(i)\,,
\end{align*}
where $0=\{\boldsymbol{r},t\}$, $i=\{\boldsymbol{r}_i,t_i\}$, $\{l\}=(0,1,...,l)$, $d(i)=d^3r_idt_i$ and with all time integrations running from $-\infty$ to $t$\,\cite{PoPReview}. Thus, any arbitrary order density response function $\mathscr{W}^{(l)}$ is formally defined by the functional derivative
\begin{align*}
\mathscr{W}^{(l)}(\{l\})&=\frac{1}{l!}\frac{\delta^{l}{n}_{\mathrm{ind}}(\boldsymbol{r},t)}{\prod_{i=1}^{l}\delta{U}_{\mathrm{ext}}(i)}\,.
\end{align*}
The arbitrary order contributions to the induced density can now be separated via the series expansion ${n}_{\mathrm{ind}}(\boldsymbol{r},t)=\sum_{l=1}^{\infty}{n}^{(l)}_{\mathrm{ind}}(\boldsymbol{r},t)$. It is apparent that they are defined by
\begin{align*}
{n}^{(l)}_{\mathrm{ind}}(\boldsymbol{r},t)&=\int\mathscr{W}^{(l)}(\{l\})\prod_{i=1}^{l}U_{\mathrm{ext}}(i)d(i)\,.
\end{align*}
Simplifications due to the system's homogeneity and stationarity are viable in the Fourier domain courtesy of the convolution theorem. The linear contribution is
\begin{align*}
{n}^{(1)}_{\mathrm{ind}}(\boldsymbol{k},\omega)&=\mathscr{W}^{(1)}(\boldsymbol{k},\omega){U}_{\mathrm{ext}}(\boldsymbol{k},\omega)\,,
\end{align*}
and the $l-$order nonlinear contribution is
\begin{align*}
{n}^{(l)}_{\mathrm{ind}}(\boldsymbol{k},\omega)&=\sum_{\substack{\sum_{1}^{l}\boldsymbol{k}_i=\boldsymbol{k}\\\sum_{1}^{l}\omega_i=\omega}}\mathscr{W}^{(l)}(\{l\}_{\mathrm{F}})\prod_{i=1}^{l}U_{\mathrm{ext}}(i_{\mathrm{F}})\,,
\end{align*}
where $i_{\mathrm{F}}=\{\boldsymbol{k}_i,\omega_i\}$ and $\{l\}_{\mathrm{F}}=(1_{\mathrm{F}},...,l_{\mathrm{F}})$. The $(l-1)$ multiplicity summation extends to all wave-vectors and to all frequencies that satisfy $\sum_{1}^{l}\boldsymbol{k}_i=\boldsymbol{k}$ and $\sum_{1}^{l}\omega_i=\omega$. As per usual, the single summation symbol implies
\begin{align*}
\sum_{\boldsymbol{k},\omega}\cdots\to\int\frac{d^3k}{(2\pi)^3}\int\frac{d\omega}{2\pi}\cdots\,.
\end{align*}
Since $U_{\mathrm{ext}}(\boldsymbol{r},t)$, ${n}_{\mathrm{ind}}(\boldsymbol{r},t)$ are real functions, the density response functions satisfy $\mathscr{W}^{(l)}(\{l\}_{\mathrm{F}})=\mathscr{W}^{(l)*}(-\{l\}_{\mathrm{F}})$. Their non-interacting limit will be denoted by $\mathscr{W}^{(l),0}$.

\textbf{Nonlinear response to harmonic perturbations}\,-- Let us consider a harmonic external potential energy perturbation $U_{\mathrm{ext}}(\boldsymbol{r},t)=2A\cos{(\boldsymbol{q}\cdot\boldsymbol{r}-\Omega{t})}$. Recall that the cosine Fourier series expansion of the even periodic function $\cos^{l}{(x)}$, with $l$ a positive integer, simply reads as
\begin{align*}
\cos^{2l}{(x)}&=\frac{c_{0,2l}}{2}+\sum_{m=1}^{l}c_{2m,2l}\cos{[(2m)x]}\,,\\
\cos^{2l+1}{(x)}&=\sum_{m=0}^{l}c_{2m+1,2l+1}\cos{[(2m+1)x]}\,,
\end{align*}
where the series coefficients are given by
\begin{align*}
c_{m,l}=\frac{1}{2^{l-1}}\binom{l}{\frac{l-m}{2}}\,.
\end{align*}
The exclusive connection between odd (even) cosine powers and odd (even) cosine harmonics is pointed out. Note that the index of the infinite series is automatically limited to the power of the cosine. In the Fourier space, this expansion implies that the nonlinear induced density of the order $2l+1$ ($2l$) involves all odd (even) harmonics from $\{\boldsymbol{q},\Omega\}$ ($2\{\boldsymbol{q},\Omega\}$) up to $(2l+1)\{\boldsymbol{q},\Omega\}$ ($2l\{\boldsymbol{q},\Omega\}$), \emph{i.e.}
\begin{align*}
&\left.{n}^{(2l)}_{\mathrm{ind}}(\boldsymbol{x})\right|_{\boldsymbol{X},A}=\sum_{m=1}^{l}\left.{n}^{(2l)}_{\mathrm{ind}}(2m\boldsymbol{X})\right|_{A}\delta_{\boldsymbol{x},2m\boldsymbol{X}}\,,\\
&\left.{n}^{(2l-1)}_{\mathrm{ind}}(\boldsymbol{x})\right|_{\boldsymbol{X},A}=\sum_{m=1}^{l}\left.{n}^{(2l-1)}_{\mathrm{ind}}[(2m-1)\boldsymbol{X}]\right|_{A}\delta_{\boldsymbol{x},(2m-1)\boldsymbol{X}}\,.
\end{align*}
In the above, we have set $\boldsymbol{x}=(\boldsymbol{k},\omega)$ $\&$ $\boldsymbol{X}=(\boldsymbol{q},\Omega)$. Note that ${n}^{(l)}_{\mathrm{ind}}(m\boldsymbol{q},m\Omega)|_{A}=A^{l}\chi^{(m,l)}_{\boldsymbol{q},\Omega}$ defines the nonlinear density response function of the order $l$ at the harmonic $m$. The total induced density is simply the sum of all nonlinear orders of the induced density, ${n}_{\mathrm{ind}}(\boldsymbol{k},\omega)|_{\boldsymbol{q},\Omega,A}=\sum_{l=1}^{\infty}{n}^{(l)}_{\mathrm{ind}}(\boldsymbol{k},\omega)|_{\boldsymbol{q},\Omega,A}$. Decomposition into even and odd nonlinear orders, use of the fact that an $m$ harmonic requires an $l\geq{m}-$order non-linearity to be excited and interchange of the summation operators lead to
\begin{align*}
\left.{n}_{\mathrm{ind}}(\boldsymbol{x})\right|_{\boldsymbol{X},A}&=\sum_{m=1}^{\infty}\sum_{l=m}^{\infty}\left.{n}^{(2l)}_{\mathrm{ind}}(2m\boldsymbol{X})\right|_{A}\delta_{\boldsymbol{x},2m\boldsymbol{X}}\nonumber\\
&+\sum_{m=1}^{\infty}\sum_{l=m}^{\infty}\left.{n}^{(2l-1)}_{\mathrm{ind}}[(2m-1)\boldsymbol{X}]\right|_{A}\delta_{\boldsymbol{x},(2m-1)\boldsymbol{X}}\,,
\end{align*}
The total induced density is also the sum of all nonlinear harmonics of the induced density, \emph{i.e}, ${n}_{\mathrm{ind}}(\boldsymbol{k},\omega)|_{\boldsymbol{q},\Omega,A}=\sum_{m=1}^{\infty}{n}_{\mathrm{ind}}(m\boldsymbol{q},m\Omega)|_{A}\delta_{\boldsymbol{k},m\boldsymbol{q}}\delta_{\omega,m\Omega}$. Decomposition into even and odd nonlinear harmonics leads to
\begin{align*}
\left.{n}_{\mathrm{ind}}(\boldsymbol{x})\right|_{\boldsymbol{X},A}&=\sum_{m=1}^{\infty}{n}_{\mathrm{ind}}(2m\boldsymbol{X})|_{A}\delta_{\boldsymbol{x},2m\boldsymbol{X}}\nonumber\\
&+\sum_{m=1}^{\infty}{n}_{\mathrm{ind}}[(2m-1)\boldsymbol{X}]|_{A}\delta_{\boldsymbol{x},(2m-1)\boldsymbol{X}}\,.
\end{align*}
From a term-by-term correspondence, the total induced density at an arbitrary odd or even harmonic is given by
\begin{align*}
&{n}_{\mathrm{ind}}(2m\boldsymbol{q},2m\Omega)|_{A}=\sum_{l=m}^{\infty}A^{2l}\chi^{(2m,2l)}_{\boldsymbol{q},\Omega}\,,\\
&{n}_{\mathrm{ind}}[(2m-1)\boldsymbol{q},(2m-1)\Omega]|_{A}=\sum_{l=m}^{\infty}A^{2l-1}\chi^{(2m-1,2l-1)}_{\boldsymbol{q},\Omega}\,.
\end{align*}
We introduce a nonlinear harmonic density response matrix $\mathcal{X}(\boldsymbol{q},\Omega)=\{\chi^{(m,l)}_{\boldsymbol{q},\Omega}\}$, whose square version is an upper triangular matrix owing to $\{\chi^{(m,l)}_{\boldsymbol{q},\Omega}\}_{m>l}\equiv0$. In view of the parity decomposition, we also define the submatrices $\mathcal{X}_{\mathrm{o}}(\boldsymbol{q},\Omega)=\{\chi^{(2m-1,2l-1)}_{\boldsymbol{q},\Omega}\}$, $\mathcal{X}_{\mathrm{e}}(\boldsymbol{q},\Omega)=\{\chi^{(2m,2l)}_{\boldsymbol{q},\Omega}\}$ and introduce the column vectors of the odd and even harmonic density perturbations, $\mathcal{P}^{\mathrm{h}}_{\mathrm{o}}(\boldsymbol{q},\Omega,A)=\{{n}_{\mathrm{ind}}[(2m-1)\boldsymbol{q},(2m-1)\Omega]|_{A}\}$, $\mathcal{P}^{\mathrm{h}}_{\mathrm{e}}(\boldsymbol{q},\Omega,A)=\{{n}_{\mathrm{ind}}(2m\boldsymbol{q},2m\Omega)|_{A}\}$, as well as the column vectors of the odd and even powers of the perturbation amplitude, $\mathcal{U}_{\mathrm{o}}(A)=\{A^{2l+1}\}$, $\mathcal{U}_{\mathrm{e}}(A)=\{A^{2l}\}$. These lead to the compact matrix forms
\begin{align*}
\mathcal{P}^{\mathrm{h}}_{\mathrm{o}}(\boldsymbol{q},\Omega,A)&=\mathcal{X}_{\mathrm{o}}(\boldsymbol{q},\Omega)\mathcal{U}_{\mathrm{o}}(A)\,,\\
\mathcal{P}^{\mathrm{h}}_{\mathrm{e}}(\boldsymbol{q},\Omega,A)&=\mathcal{X}_{\mathrm{e}}(\boldsymbol{q},\Omega)\mathcal{U}_{\mathrm{e}}(A)\,.
\end{align*}

\textbf{Connection between harmonic and general density response functions}\,-- The connection between the general density response $\mathscr{W}^{(l)}(\boldsymbol{k}_1,\omega_1;\cdots;\boldsymbol{k}_l,\omega_l)$ and the harmonic density response $\chi^{(m,l)}_{\boldsymbol{q},\Omega}$, for an arbitrary nonlinear order $l$ and an arbitrary harmonic $m$, will now be elucidated. On the basis of the above cosine Fourier series and after some cumbersome $\delta$-function algebra, the following relation can be established;
\begin{align*}
\chi^{(m,l)}_{\boldsymbol{q},\Omega}&=\mathcal{S}\mathscr{W}^{(l)}(\underbrace{\boldsymbol{q},\Omega;\cdots;\boldsymbol{q},\Omega}_{\frac{l+m}{2}\,\mathrm{terms}};\underbrace{-\boldsymbol{q},-\Omega;\cdots;-\boldsymbol{q},-\Omega}_{\frac{l-m}{2}\,\mathrm{terms}})
\end{align*}
where $\mathcal{S}$ stands for the sum of the different combinations of the $(l-m)/2$ negative argument pairs from the set of $l$ argument pairs. Note that the number of contributions is $\binom{l}{[l-m]/2}$, see also the $c_{m,l}$ coefficient of the cosine Fourier series. For instance, $\chi^{(1,5)}_{\boldsymbol{q},\Omega}$ features ten $\mathscr{W}^{(l)}$ adders, while $\chi^{(3,5)}_{\boldsymbol{q},\Omega}$ features five $\mathscr{W}^{(l)}$ adders. Considering the compact matrix representation, the harmonic responses $\chi^{(l,l)}_{\boldsymbol{q},\Omega}$ can be coined as diagonal density responses and the harmonic responses $\chi^{(m,l)}_{\boldsymbol{q},\Omega}$ with $m<l$ can be coined as off-diagonal density responses. In what follows, we shall focus on the diagonal responses, since they represent the dominant contribution to any harmonic within the weak turbulence limit. For the diagonal density responses, we simply have
\begin{align*}
\chi^{(l,l)}_{\boldsymbol{q},\Omega}&=\mathscr{W}^{(l)}(\boldsymbol{q},\Omega;\cdots;\boldsymbol{q},\Omega)\,.
\end{align*}

\textbf{Nonlinear non-interacting density responses}\,-- Quantum kinetic theory in the Wigner representation\,\cite{BonitzBok} constitutes the most straightforward method for the calculation of nonlinear ideal density responses in complete analogy with the classical kinetic theory\,\cite{UspekhiR1}. In the non-interacting limit, the evolution of the one-particle distribution function $f(\boldsymbol{r},\boldsymbol{p},t)$, \emph{i.e.}, the Wigner quasiprobability distribution\,\cite{UspekhiR2}, is described by the $s=1$ member of the quantum Wigner-BBGKY hierarchy, that automatically becomes decoupled in absence of interactions\,\cite{BonitzBok},
\begin{align*}
&\left\{\frac{\partial}{\partial{t}}+\boldsymbol{v}\cdot\frac{\partial}{\partial\boldsymbol{r}}\right\}f(\boldsymbol{r},\boldsymbol{p},t)=\frac{\imath}{\hbar}\int\,\frac{d^3\lambda{d}^3\bar{p}}{(2\pi)^3}\exp{\left[\imath\left(\boldsymbol{p}-\boldsymbol{\bar{p}}\right)\cdot\boldsymbol{\lambda}\right]}\nonumber\\&\,\,\,\left[U_{\mathrm{ext}}\left(\boldsymbol{r}+\frac{\hbar}{2}\boldsymbol{\lambda},t\right)-U_{\mathrm{ext}}\left(\boldsymbol{r}-\frac{\hbar}{2}\boldsymbol{\lambda},t\right)\right]f(\boldsymbol{r},\boldsymbol{\bar{p}},t).
\end{align*}
Let us denote the unperturbed distribution function with $f_0(\boldsymbol{p})$ and its perturbation due to the external potential energy perturbation with $f^{\mathrm{ind}}(\boldsymbol{r},\boldsymbol{p},t)$. It is evident that $f(\boldsymbol{r},\boldsymbol{p},t)=f_0(\boldsymbol{p})+f^{\mathrm{ind}}(\boldsymbol{r},\boldsymbol{p},t)$. Assuming the adiabatic switching of the perturbation in the infinite past when the homogeneous medium was in thermodynamic equilibrium, the Fourier transformed kinetic equation reads
\begin{align*}
&f^{\mathrm{ind}}_{\boldsymbol{p},\boldsymbol{k},\omega}=-\frac{1}{\hbar}\left[\frac{f_0\left(\boldsymbol{p}+\frac{\hbar}{2}\boldsymbol{k}\right)-f_0\left(\boldsymbol{p}-\frac{\hbar}{2}\boldsymbol{k}\right)}{\omega-\boldsymbol{k}\cdot\boldsymbol{v}+\imath0}\right]{U}^{\mathrm{ext}}_{\boldsymbol{k},\omega}-\nonumber\\&\,\,\,\frac{1}{\hbar}\int\,\frac{d\omega^{\prime}d^3k^{\prime}}{(2\pi)^4}\left[\frac{f^{\mathrm{ind}}_{\boldsymbol{p}+\frac{\hbar}{2}\boldsymbol{k}^{\prime},\boldsymbol{k}-\boldsymbol{k}^{\prime},\omega-\omega^{\prime}}-f^{\mathrm{ind}}_{\boldsymbol{p}-\frac{\hbar}{2}\boldsymbol{k}^{\prime},\boldsymbol{k}-\boldsymbol{k}^{\prime},\omega-\omega^{\prime}}}{\omega-\boldsymbol{k}\cdot\boldsymbol{v}+\imath0}\right]{U}^{\mathrm{ext}}_{\boldsymbol{k}^{\prime},\omega^{\prime}}\,.
\end{align*}
Setting $\boldsymbol{p}=\hbar\boldsymbol{q}$, dropping $\hbar$ from the distribution function arguments and introducing $\epsilon_{\boldsymbol{q}}=[\hbar^2/(2m)]q^2$ for the non-interacting gas, leads to
\begin{align*}
&f^{\mathrm{ind}}_{\boldsymbol{q},\boldsymbol{k},\omega}=-\left[\frac{f_0\left(\boldsymbol{q}+\frac{1}{2}\boldsymbol{k}\right)-f_0\left(\boldsymbol{q}-\frac{1}{2}\boldsymbol{k}\right)}{\hbar\omega-\epsilon_{\boldsymbol{q}+\frac{1}{2}\boldsymbol{k}}+\epsilon_{\boldsymbol{q}-\frac{1}{2}\boldsymbol{k}}+\imath0}\right]{U}^{\mathrm{ext}}_{\boldsymbol{k},\omega}-\nonumber\\&\,\,\,\int\,\frac{d\omega^{\prime}d^3k^{\prime}}{(2\pi)^4}\left[\frac{f^{\mathrm{ind}}_{\boldsymbol{q}+\frac{1}{2}\boldsymbol{k}^{\prime},\boldsymbol{k}-\boldsymbol{k}^{\prime},\omega-\omega^{\prime}}-f^{\mathrm{ind}}_{\boldsymbol{q}-\frac{1}{2}\boldsymbol{k}^{\prime},\boldsymbol{k}-\boldsymbol{k}^{\prime},\omega-\omega^{\prime}}}{\hbar\omega-\epsilon_{\boldsymbol{q}+\frac{1}{2}\boldsymbol{k}}+\epsilon_{\boldsymbol{q}-\frac{1}{2}\boldsymbol{k}}+\imath0}\right]{U}^{\mathrm{ext}}_{\boldsymbol{k}^{\prime},\omega^{\prime}}\,.\nonumber
\end{align*}
The first RHS term is the standard linear contribution, whereas the second RHS term is the nonlinear contribution that stems from the Fourier multiplication property. The nonlinear equation can be iterated, \emph{i.e.}, the distribution function perturbation can be expanded according to $f^{\mathrm{ind}}_{\boldsymbol{q},\boldsymbol{k},\omega}=\sum_{l=1}^{\infty}f^{\mathrm{ind},(l)}_{\boldsymbol{q},\boldsymbol{k},\omega}$. This procedure is reminiscent of the time-dependent perturbation theory of quantum mechanics\,\cite{Quantpert} and the weak turbulence theory of plasma physics\,\cite{Weakturbu}. Ultimately, it yields the recurrence relation
\begin{multline*}
f^{\mathrm{ind},(l)}_{\boldsymbol{q},\boldsymbol{k},\omega}=\\\int\,\frac{d\omega^{\prime}d^3k^{\prime}}{(2\pi)^4}\frac{f^{\mathrm{ind},(l-1)}_{\boldsymbol{q}-\frac{1}{2}\boldsymbol{k}^{\prime},\boldsymbol{k}-\boldsymbol{k}^{\prime},\omega-\omega^{\prime}}-f^{\mathrm{ind},(l-1)}_{\boldsymbol{q}+\frac{1}{2}\boldsymbol{k}^{\prime},\boldsymbol{k}-\boldsymbol{k}^{\prime},\omega-\omega^{\prime}}}{\hbar\omega-\epsilon_{\boldsymbol{q}+\frac{1}{2}\boldsymbol{k}}+\epsilon_{\boldsymbol{q}-\frac{1}{2}\boldsymbol{k}}+\imath0}{U}^{\mathrm{ext}}_{\boldsymbol{k}^{\prime},\omega^{\prime}}\,,
\end{multline*}
where the linear contribution is given by
\begin{align*}
f^{\mathrm{ind},(1)}_{\boldsymbol{q},\boldsymbol{k},\omega}&=\frac{f_0\left(\boldsymbol{q}-\frac{1}{2}\boldsymbol{k}\right)-f_0\left(\boldsymbol{q}+\frac{1}{2}\boldsymbol{k}\right)}{\hbar\omega-\epsilon_{\boldsymbol{q}+\frac{1}{2}\boldsymbol{k}}+\epsilon_{\boldsymbol{q}-\frac{1}{2}\boldsymbol{k}}+\imath0}{U}^{\mathrm{ext}}_{\boldsymbol{k},\omega}\,.
\end{align*}
Successive application of the recurrence relation, utilization of the $n_{\mathrm{ind}}^{(l)}(\boldsymbol{k},\omega)=2\int\,[d^3q/(2\pi)^3]f^{\mathrm{ind},(l)}_{\boldsymbol{q},\boldsymbol{k},\omega}$ normalization condition, change of variables to unify the argument of the Fermi-Dirac distribution and $l$-order functional differentiation with respect to the external potential energy, yield analytical expressions for the Fourier transformed $\mathscr{W}^{(l),0}$. Thus, cumbersome $l-$order closed form expressions can be derived, but they have no practical use. Nevertheless, it is very important to point out that $\mathscr{W}^{(l),0}$ comprises $2^l$ alternating sign contributions of the form
\begin{align*}
\int\frac{d^3q}{(2\pi)^3}\frac{2f_0(\boldsymbol{q})}{\prod_{i=1}^{l}\left[\hbar\sum_{j=1}^{i}z_j-\epsilon_{\boldsymbol{q}+\boldsymbol{q}_i}+\epsilon_{\boldsymbol{q}+\boldsymbol{q}_i-\sum_{j=1}^{i}\boldsymbol{k}_j}\right]}\,,
\end{align*}
where analytic continuation of the $l$-frequencies has been performed in the complex coordinate space, $\omega_i+\imath0\to{z}_i$, and with $\boldsymbol{q}_i=\sum_{j=1}^{i}c_j\boldsymbol{k}_j$, $c_j\in\{0,\pm1\}$. In a similar way, the diagonal non-interacting density response $\chi^{(l,l)}_{0}(\boldsymbol{q},\Omega)$ comprises $2^l$ alternating sign contributions of the form
\begin{align*}
\frac{1}{l!}\int\frac{d^3k}{(2\pi)^3}\frac{2f_0(\boldsymbol{k})}{\prod_{i=1}^{l}\left[\hbar{Z}-\frac{\epsilon_{\boldsymbol{k}+d_i\boldsymbol{q}}-\epsilon_{\boldsymbol{k}+d_i\boldsymbol{q}-i\boldsymbol{q}}}{i}\right]}\,,
\end{align*}
where $d_i=\{0,\cdots,i\}$, $\Omega+\imath0\to{Z}$ and the dummy integration variable has been changed to $\boldsymbol{k}$ to avoid notational conflict with the perturbation wavenumber $\boldsymbol{q}$.

\begin{figure*}
\includegraphics[width=0.333\textwidth]{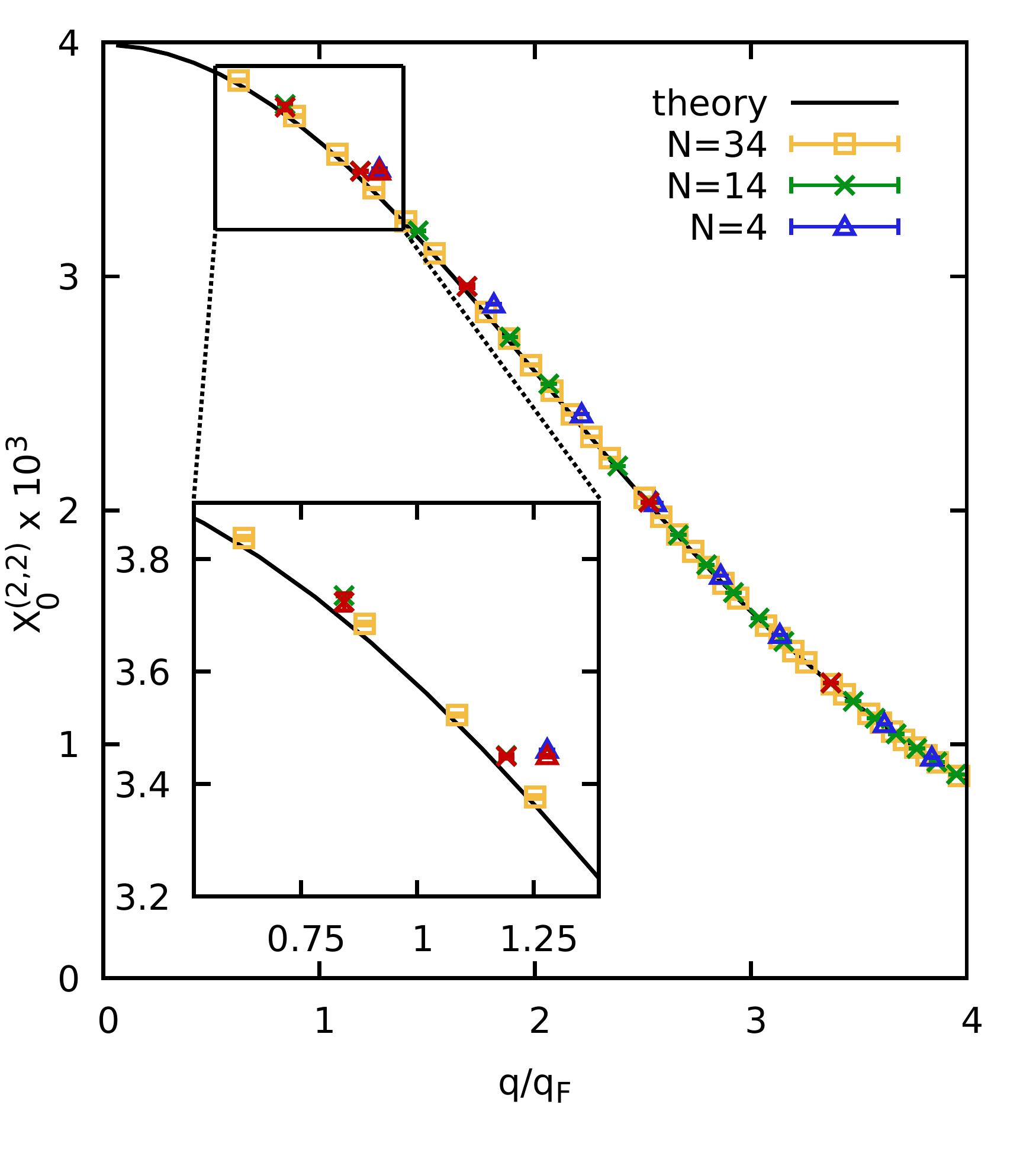}\includegraphics[width=0.333\textwidth]{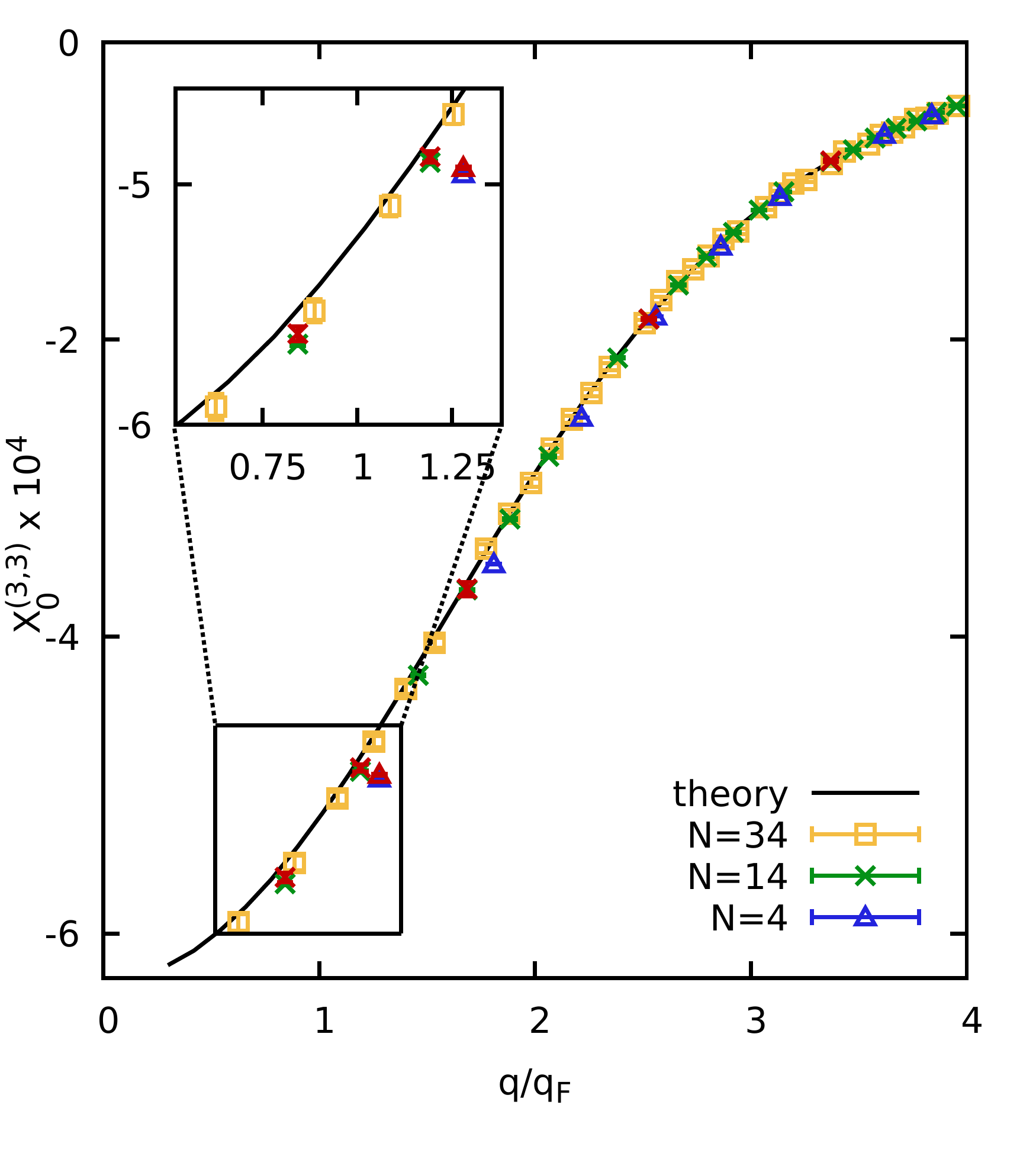}\includegraphics[width=0.333\textwidth]{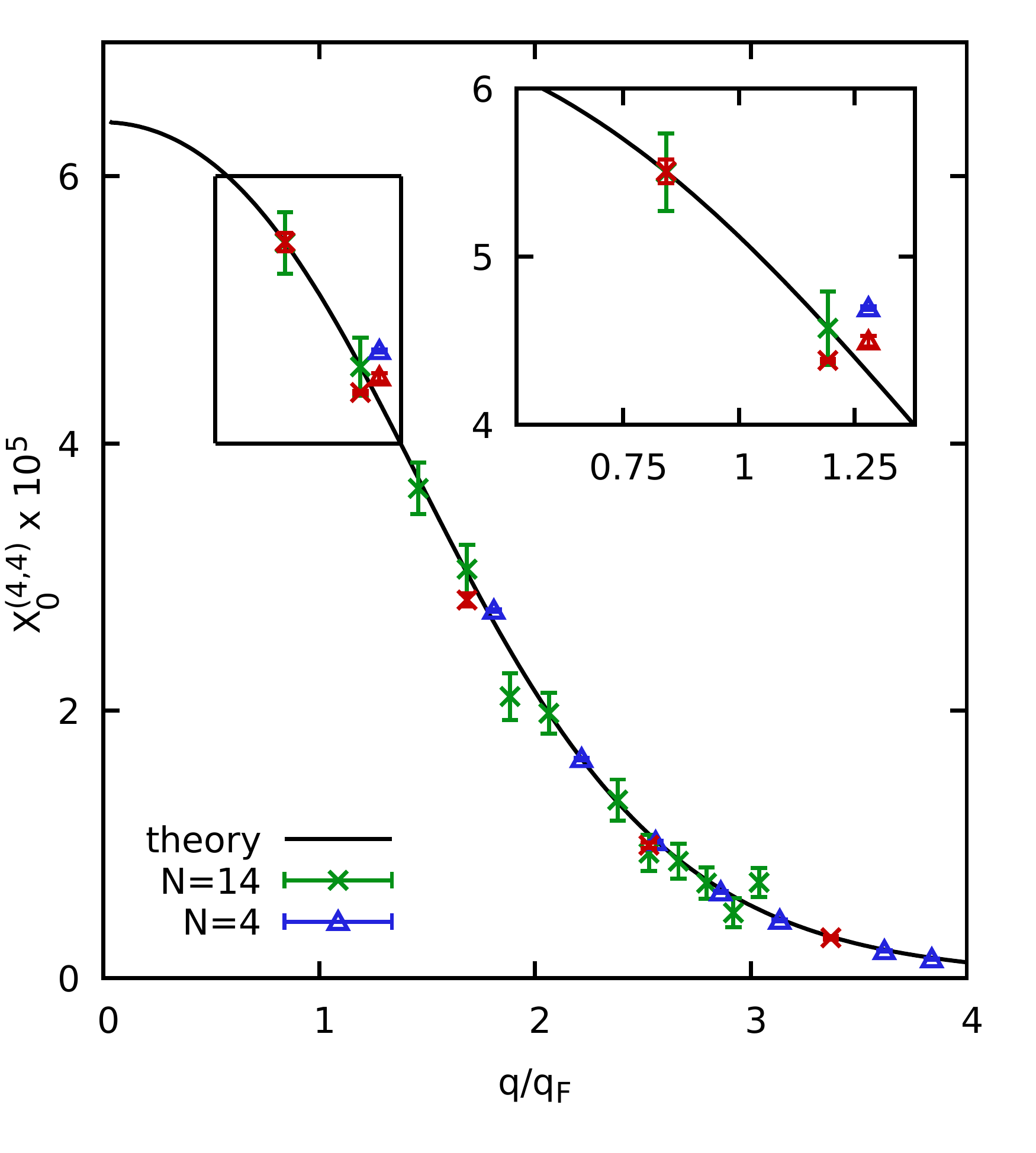}\vspace*{-0.41cm}
\caption{\label{fig:UEG}
The nonlinear static density response functions of the ideal Fermi gas at $r_s=2$, $\Theta=4$ at the second, third, and fourth harmonics (from left to right). Solid black line: exact recursion relation as in the main text; yellow squares, green crosses, blue triangles: \emph{ab initio} PIMC results for different system size as computed from integrated higher-order imaginary-time correlation functions~\cite{Dornheim4}; corresponding red symbols: results from PIMC simulations of the harmonically perturbed system~\cite{Dornheim2,PoPReview}.
}
\end{figure*}

\textbf{Canonical form of the diagonal non-interacting density response}\,\cite{suppleme1}--  The observation that each of the $2^l$ contributions to $\chi^{(l,l)}_{0}(\boldsymbol{q},Z)$ can be considered as a rational function of the complex frequency allows the use of partial fraction decomposition for the automatically factorized denominator. It needs to be noted that, starting from $l\geq3$, some of the contributions contain factors which satisfy $j[\epsilon_{\boldsymbol{k}+d_i\boldsymbol{q}}-\epsilon_{\boldsymbol{k}+d_i\boldsymbol{q}-i\boldsymbol{q}}]=i[\epsilon_{\boldsymbol{k}+d_j\boldsymbol{q}}-\epsilon_{\boldsymbol{k}+d_j\boldsymbol{q}-j\boldsymbol{q}}]$ for $i,j\in\{1,\cdots,l\}$. This implies that the denominator can have roots of varying multiplicity, which significantly complicates the decomposition. Brute force calculations have been carried out for $l=2,3,4,5$. These calculations become exceedingly cumbersome; for $l=5$, there are $32$ standard form contributions featuring quintic polynomials of $Z$ leading to $160$ adders in the response prior to simplifications. Nevertheless, for $l=2,3,4,5$, all the partial fractions that feature multiple roots cancel out exactly and all the surviving partial fractions that feature simple roots can be combined in the form of a Lindhard ideal density response $\chi^{(1,1)}_{0}(\boldsymbol{q},Z)\equiv\chi_{0}(\boldsymbol{q},Z)$ evaluated at all the possible $m\leq{l}$ harmonics. The compact expressions, normalized according to $\widetilde{\chi}^{(l,l)}_{0}(\boldsymbol{x},Z)=\chi^{(l,l)}_{0}(\boldsymbol{x},Z)/(n\beta^l)$ where $\boldsymbol{x}=\boldsymbol{q}/q_{\mathrm{F}}$ with $n$ the density, $\beta$ the inverse temperature and $q_{\mathrm{F}}$ the Fermi wave-vector, for the quadratic, cubic, quartic and quintic ideal density responses read as
\begin{align}\label{eq:eq}
&\widetilde{\chi}^{(2,2)}_{0}(\boldsymbol{x},Z)=\frac{1}{1!}\frac{1}{2!}\frac{\Theta}{x^2}\left[2\widetilde{\chi}_0(2\boldsymbol{x},2Z)-2\widetilde{\chi}_0(\boldsymbol{x},Z)\right]\,,\\
&\widetilde{\chi}^{(3,3)}_{0}(\boldsymbol{x},Z)=\frac{1}{2!}\frac{1}{3!}\frac{\Theta^2}{x^4}\left[3\widetilde{\chi}_0(3\boldsymbol{x},3Z)-8\widetilde{\chi}_0(2\boldsymbol{x},2Z)\right.\nonumber\\&\quad\quad\quad\quad\quad\quad\left.+5\widetilde{\chi}_0(\boldsymbol{x},Z)\right]\,,\\
&\widetilde{\chi}^{(4,4)}_{0}(\boldsymbol{x},Z)=\frac{1}{3!}\frac{1}{4!}\frac{\Theta^3}{x^6}\left[4\widetilde{\chi}_0(4\boldsymbol{x},4Z)-18\widetilde{\chi}_0(3\boldsymbol{x},3Z)\right.\nonumber\\&\quad\quad\quad\quad\quad\quad\left.+28\widetilde{\chi}_0(2\boldsymbol{x},2Z)-14\widetilde{\chi}_0(\boldsymbol{x},Z)\right]\,,\\
&\widetilde{\chi}^{(5,5)}_{0}(\boldsymbol{x},Z)=\frac{1}{4!}\frac{1}{5!}\frac{\Theta^4}{x^8}\left[5\widetilde{\chi}_0(5\boldsymbol{x},5Z)-32\widetilde{\chi}_0(4\boldsymbol{x},4Z)\right.\nonumber\\&\quad\left.+81\widetilde{\chi}_0(3\boldsymbol{x},3Z)-96\widetilde{\chi}_0(2\boldsymbol{x},2Z)+42\widetilde{\chi}_0(\boldsymbol{x},Z)\right].
\end{align}
The first two expressions were derived by Mikhailov who solved the non-interacting quantum kinetic equation for the density matrix with perturbation techniques\,\cite{Mikhailo1,Mikhailo2}. Based on the above exact results, we make the conjecture that the $l-$order ideal density response at the $l$ harmonic can be expressed as the weighted sum of Lindhard density responses evaluated from the fundamental harmonic up to the $l-$harmonic. The canonical form Ansatz reads as
\begin{align*}
&\widetilde{\chi}^{(l,l)}_{0}(\boldsymbol{x},Z)\propto\sum_{m=1}^{l}g_{m,l}(x)\widetilde{\chi}_0(m\boldsymbol{x},mZ)\nonumber\,.
\end{align*}
A simple recipe for the general computation of the coefficients $g_{m,l}(x)$ can be constructed based on the following observations. \textbf{(1)} Regardless of $l$, the origin of the highest $m=l$ harmonic term can be traced back to the only contribution that features a factor $\hbar{Z}-(\epsilon_{\boldsymbol{k}}-\epsilon_{\boldsymbol{k}-l\boldsymbol{q}})/l$ and the only contribution that features a factor $\hbar{Z}-(\epsilon_{\boldsymbol{k}+l\boldsymbol{q}}-\epsilon_{\boldsymbol{k}})/l$ in the denominator. This yields a general form for $g_{l,l}(x)$,
\begin{align*}
g_{l,l}(x)=\frac{1}{[(l-1)!]^2}\frac{\Theta^{l-1}}{x^{2l-2}}\,.\nonumber
\end{align*}
The factorization of the function $a(x)=g_{l,l}(x)/l$ leads to integer canonical coefficients ${a}_{m,l}=g_{m,l}(x)/a(x)$. Thus, the canonical form becomes
\begin{align}
&\widetilde{\chi}^{(l,l)}_{0}(\boldsymbol{x},Z)=\frac{1}{(l-1)!}\frac{1}{l!}\frac{\Theta^{l-1}}{x^{2l-2}}\left[l\widetilde{\chi}_0(l\boldsymbol{x},lZ)+\sum_{m=1}^{l-1}a_{m,l}\times\right.\nonumber\\&\quad\quad\quad\quad\quad\quad\quad\quad\quad\quad\quad\quad\quad\left.\widetilde{\chi}_0(m\boldsymbol{x},mZ)\right]\,.\label{eq:canonicalexpansion}
\end{align}
\textbf{(2)} From the above expressions, it can be verified that the leading-order term of the long-wavelength limit $(x\to0)$ of the diagonal ideal static $(Z\equiv0)$ density responses is constant. Considering the validity of this observation for $l=2,3,4,5$, we assume that it also holds for arbitrary $l$ values. Given the presence of the $1/x^{2l-2}$ pre-factor that should be cancelled exactly and given the fact that the Taylor expansion of the static Lindhard density response contains only even powers of $x$, it is apparent that this constraint will lead to $(l-1)$ linear equations that can be solved for the remaining $(l-1)$ unknown coefficients. In particular, the static Lindhard density response is
\begin{align*}
\widetilde{\chi}_{0}(\boldsymbol{x})&=-\frac{3}{2}\frac{\Theta}{x}\int_0^{\infty}yf_0(y;\Theta)\ln{\left|\frac{x+2y}{x-2y}\right|}dy\nonumber\,.
\end{align*}
Taylor expansion of the logarithms for $x/y\ll1$ yields
\begin{align*}
\widetilde{\chi}_{0}(\boldsymbol{x})&=-\frac{3}{2}\Theta\int_0^{\infty}dyf_0(y;\Theta)\displaystyle\sum_{\nu=0}^{\infty}\frac{1}{2\nu+1}\left(\frac{x}{2y}\right)^{2\nu}\,.
\end{align*}
Substitution in the canonical form leads to
\begin{align*}
&\lim_{x\to0}\widetilde{\chi}^{(l,l)}_{0}(\boldsymbol{x})=-\frac{3}{2}\frac{1}{(l-1)!l!}\Theta^{l}\lim_{x\to0}\displaystyle\sum_{\nu=0}^{\nu=l-1}\int_0^{\infty}dy\frac{f_0(y;\Theta)}{y^{2\nu}}\nonumber\\&\qquad\,\,\,\left\{\frac{1}{2\nu+1}\frac{l^{2\nu+1}}{2^{2\nu}}+\displaystyle\sum_{m=1}^{l-1}\frac{1}{2\nu+1}\frac{a_{m,l}m^{2\nu}}{2^{2\nu}}\right\}x^{2\nu-2l+2}\,.\nonumber
\end{align*}
For the constant $\nu=l-1$ term to survive and the long wavelength limit to exist, the bracketed term must vanish $\forall\nu=0,1,\cdots,{l}-2$. This leads to the desired $(l-1)\times(l-1)$ system of linear equations for the coefficients $a_{m,l}$:
\begin{align}
&\displaystyle\sum_{m=1}^{l-1}a_{m,l}m^{2\nu}=-l^{2\nu+1},\,\,\mathrm{for}\,\,\nu=0,1,...l-2\,.\label{eq:coefficients}
\end{align}
Computations up to $l=100$ have revealed that the $a_{m,l}$ coefficients are always integers and that the $a_{m,l}$ sign is alternating $\forall{l}$. We emphasize that even though the $a_{m,l}$ coefficients are calculated via the static long-wavelength limit, the canonical form is valid for any wave-number and any frequency. The compact expression for the sextic ideal density response is given below as an example;
\begin{align}
&\widetilde{\chi}^{(6,6)}_{0}(\boldsymbol{x},Z)=\frac{1}{5!6!}\frac{\Theta^5}{x^{10}}\left[6\widetilde{\chi}_0(6\boldsymbol{x},6Z)-50\widetilde{\chi}_0(5\boldsymbol{x},5Z)\right.\nonumber\\&\qquad\,\,\,+176\widetilde{\chi}_0(4\boldsymbol{x},4Z)-330\widetilde{\chi}_0(3\boldsymbol{x},3Z)+330\widetilde{\chi}_0(2\boldsymbol{x},2Z)\nonumber\\&\qquad\,\,\,\left.-132\widetilde{\chi}_0(\boldsymbol{x},Z)\right]\,.
\end{align}

\textbf{Results}\,-- In the static case, the higher-order recursion relations are benchmarked against new \emph{ab initio} PIMC simulations of the non-interacting Fermi gas. Comparisons are shown in Fig.\ref{fig:UEG} for the quadratic (left), cubic (center), and quartic (right) ideal static density response. The solid black lines have been computed from the recursion relations. The yellow squares, green crosses, and blue triangles correspond to the PIMC results for different system size $N$ computed via the integration of higher-order imaginary-time correlation functions~\cite{Dornheim2}. It is evident that the $N$-dependence is small and restricted to the long wavelength limit; consistent with previous investigations of similar wavenumber-resolved properties~\cite{DPRL_2016,DFSC_2021}. The respective red symbols have been obtained directly from PIMC simulations of the harmonically perturbed system~\cite{Dornheim2,PoPReview}. As expected, they are in good agreement with the imaginary-time based data and the analytical black curve valid in the thermodynamic limit ($N\to\infty$).

\begin{figure}
\includegraphics[width=0.4\textwidth]{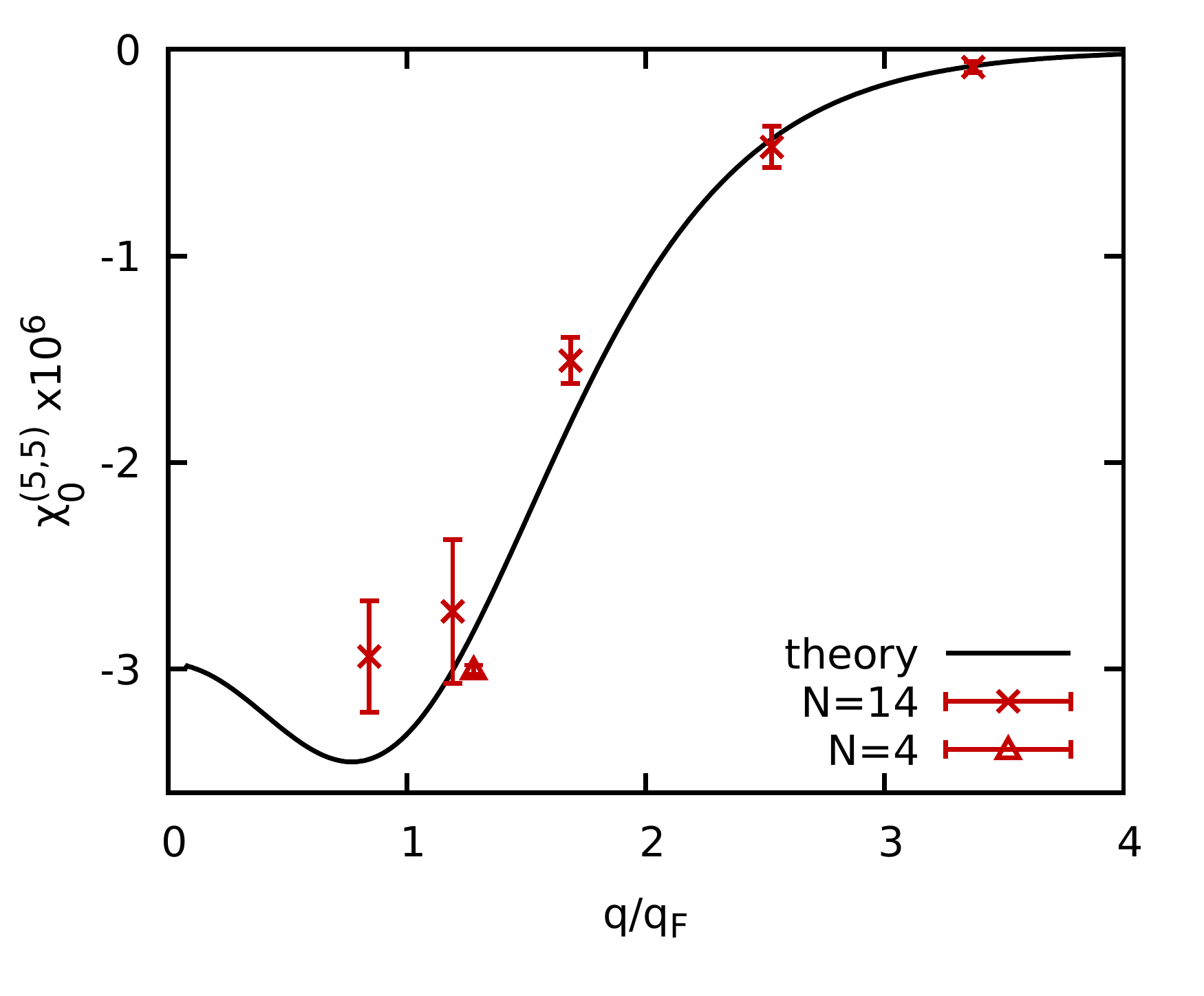}\vspace*{-0.4cm}
\caption{\label{fig:UEG_quintic} The ideal static density response function at the fifth harmonic ($r_s=2$, $\Theta=4$). Solid black line: exact recursion relation as in the main text; red symbols: direct PIMC results for $N=4$ (triangles) and $N=14$ (crosses).
}
\end{figure}

In Fig.\ref{fig:UEG_quintic}, comparison is presented for the quintic ideal static density response. The higher-order imaginary-time technique would now require the PIMC estimation of a correlator of six density operators with different time arguments, which is cumbersome in practice. In addition, the small magnitude of the quintic density response limits accurate results from the direct perturbation technique to $N=14$ and $N=4$. Still, the results are in excellent agreement with the exact recursion relation within the given level of accuracy over the full wavenumber-range. Note the emergence of an extremum not present at lower orders. For $\Theta=4$, Eqs.(\ref{eq:canonicalexpansion},\ref{eq:coefficients}) yield shallow or deep extrema at $q\simeq{q}_{\mathrm{F}}$ for $l\geq5$, at least up to the decic level.

\begin{figure}
\includegraphics[width=0.4\textwidth]{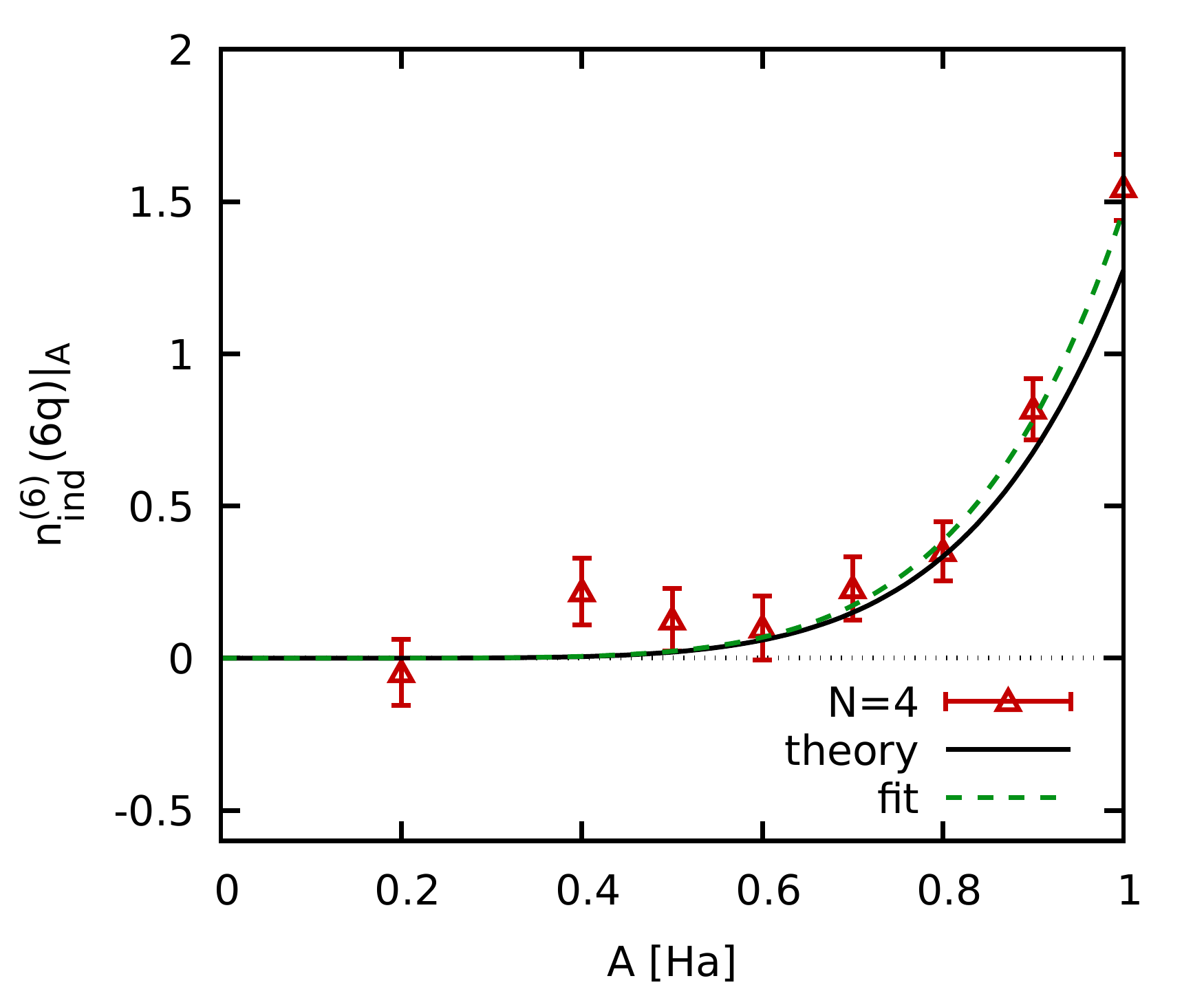}\vspace*{-0.4cm}
\caption{\label{fig:UEG_sextic} The ideal static induced density at the sixth harmonic for $q/q_\textnormal{F}=1.28$ ($r_s=2$, $\Theta=4$). Red triangles: direct PIMC results for $N=4$; dashed green line: $n_{\mathrm{ind}}^{(6)}(6q)|_{A}=\chi_0^{(6,6)}(q)A^6$ fit with $\chi_0^{(6,6)}(q)$ treated as the fit parameter; solid black line: replacement of the fit parameter with the theoretical $\chi_0^{(6,6)}(q)$ computed from the recursion relation.
}
\end{figure}

Accurate estimation of the sextic ideal static density response from direct PIMC simulations is even more challenging compared to the quintic case, making a wavenumber scan unfeasible. In Fig.\ref{fig:UEG_sextic}, the static induced density at the sixth harmonic $n_{\mathrm{ind}}^{(6)}(6q)|_{A}$ is shown as function of the perturbation amplitude $A$ for $q/q_\textnormal{F}=1.28$, with the red triangles depicting $N=4$ PIMC data. The dashed green line is obtained from a $n_{\mathrm{ind}}^{(6)}(6q)|_{A}=\chi_0^{(6,6)}(q)A^6$ fit over $A\in[0,0.8]$  with $\chi_0^{(6,6)}(q)$ being the free parameter; the solid black line corresponds to the same functional form, but with $\chi_0^{(6,6)}(q)$ computed from the theoretical recursion relation. Both curves are in good agreement with the direct PIMC results within the given error bars, and agree well with each other. This nicely substantiates the validity of the inferred general recursion relation.

\textbf{Discussion}\,-- The canonical form of the diagonal ideal density responses constitutes the arbitrary-order generalization of the ideal linear (Lindhard) density response. It completes the earlier work of Mikhailov, who carried out the second- and third-order generalizations~\cite{Mikhailo2}. It is also connected with an earlier work of Tanaka, who derived an elegant expression for arbitrary order correlation functions of the non-interacting Fermi gas at the ground state~\cite{TanakaShi}. Being motivated by warm dense matter applications, the present exposition was based on the three-dimensional paramagnetic ideal electron gas. Nevertheless, the canonical expression is a very general result. In fact, the derivation includes neither a wavenumber-space integration nor a substitution of the equilibrium distribution. This implies that the canonical form is valid, albeit with different normalizations, for: (i) any system dimensionality, (ii) any spin polarization, (iii) any quantum statistics including fractional exclusion statistics, (iv) the classical limit where the Lindhard response collapses to the Vlasov response, (v) certain nonequilibrium stationary states. Future work will focus on the offdiagonal ideal density responses, for which the perturbative treatment of the nonlinear response theory leads to divergences regardless of the harmonic order and the nonlinearity order. These concern the general response function, \emph{i.e.} the building block of the harmonic response function via the combinatorial expression, and, thus, could be removable.

\emph{Acknowledgments.} This work was partially supported by the Center for Advanced Systems Understanding (CASUS) which is financed by Germany's Federal Ministry of Education and Research (BMBF) and the Saxon state government out of the state budget that is approved by the Saxon State Parliament. The PIMC simulations were carried out at the Norddeutscher Verbund f\"ur Hoch- und H\"ochstleistungsrechnen (HLRN) under grant shp00026, and on a Bull Cluster at the Center for Information Services and High Performance Computing (ZIH) at Technische Universit\"at Dresden.

\end{document}